\documentclass[10pt,a4paper]{revtex4}
 
\usepackage{hyperref}
\usepackage{amsmath,amsfonts}
\usepackage[dvips]{graphics}
\usepackage[dvips]{graphicx}
\usepackage{subfigure}
\usepackage{epsfig}
 
\begin{document}
\title{\large Leading-order corrections to charged rotating AdS black holes thermodynamics}
\author{Sudhaker Upadhyay}
  \email{sudhakerupadhyay@gmail.com; sudhaker@associates.iucaa.in}
 
\affiliation{Department of Physics, K.L.S. College, Magadh University, Nawada-805110, India}
\affiliation{Visiting Associate,  Inter-University Centre for Astronomy
and Astrophysics (IUCAA) Pune, Maharashtra-411007}

\begin{abstract}In this paper, we consider a charged rotating AdS black holes in four dimensions and study the effects of leading-order thermal corrections  on the thermodynamics of such system explicitly. The first-order 
corrected thermodynamical quantities  also satisfy the first-law of thermodynamics of the black holes. The holographic duality between the charged
 rotating AdS black holes and Van der Waals fluid is also emphasized through the $P-v$ diagram. Finally, we study the effects of the leading-order thermal corrections on the stability of the charged rotating  black holes. 
 \end{abstract}

\maketitle

\section{Overview and motivation}
In the last few decades, black hole thermodynamics has been playing an important role
 in order  to understand quantum gravity. 
 It is started with Hawking and Page who found that the black hole solutions in asymptotically AdS space have thermodynamic
properties including a characteristic temperature and an intrinsic entropy
proportional to the area of the event horizon \cite{haw}.
  In this connection, assuming the black hole as a thermodynamic system in equilibrium at Hawking temperature,
  its small statistical fluctuations around equilibrium is systematically studied diversely.
   These fluctuations give rise to a
non-trivial multiplicative factor to the expression for the density of states.  This leads to
a correction to the entropy of black hole which is logarithmic,   determined macroscopically from the massless
particle spectrum, at the leading order \cite{l1}. 
These corrections are computed 
microscopically as well using a stringy embedding of the Kerr/CFT correspondence
and found perfect agreement \cite{stro}.
The first-order quantum correction to the  entropy  is also computed  
by examining the logarithmic corrections to the Cardy formula \cite{card}.
This confirms that logarithmic corrections to
the entropy originated due to the statistical fluctuations around thermal equilibrium of the black hole.

The  effects of  thermal fluctuations on the thermodynamics of various black holes have been studied. For instance, more recently,  the effect
of thermal fluctuations on the thermodynamics of a massive gravity black hole in   AdS space
is studied \cite{sud}. From the  critical points and stability analysis, it has been shown that the presence of logarithmic correction is necessary to have stable
phase and critical point. On the other hand,  the Van der Waals phase
transitions  for the massive gravity  black holes is also studied and found that massive gravity theories can exhibit
strikingly different thermodynamic behavior compared to that of Einstein gravity, and that the mass of the
graviton can generate a range of new phase transitions   that are 
otherwise forbidden \cite{hend}. 
 The quantum gravity effects on Ho\v rava-Lifshitz black hole 
is also investigated  and discussed such effects on the black hole stability \cite{sud1}.
By comparing the corrected entropy to  the energy  fluctuations   and  a conformal field theory,  the corrections to the thermodynamics and  the phase transition of  charged dilatonic black Saturn   are studied \cite{mir0}.
Recently, the  thermodynamical properties of recently proposed Schwarzschild-Beltrami-de Sitter black hole with the higher order corrected entropy  are also analysed \cite{sud2}. Here,   first and second laws of thermodynamics are  established and found that presence of higher order corrections, stem  from thermal fluctuations, may remove some instabilities of the black hole.  The logarithm-corrected entropy and  thermodynamics of a dyonic charged 
AdS  black hole is studied and  holographic duality of a
van der Waals fluid is established \cite{beh}. The effects of first-order correction
to the modified Hayward black hole have been studied,
where it has been found that corrections reduce the
pressure and internal energy of the Hayward black hole \cite{mirf}.
Recently, based on the Jacobson formalism, a method is proposed to test the effects of quantum fluctuations on black holes through the thermal fluctuations on dumb holes \cite{mir1}
We are interested to study the effects of leading-order correction on the stability and thermodynamics of four dimensional charged rotating AdS black hole.

 In this paper, we consider the charged rotating black holes as thermal system  and try to analyse the effects of  thermal fluctuations on the thermodynamics of such black hole. We fist compute the  Hawking temperature and leading-order correction to the entropy of the system originated due to the
 thermal fluctuations which is logarithmic in nature. 
 We see that entropy is an increasing
function for large black holes when system is in equilibrium. However, due to thermal fluctuation, the
entropy does not exist for smaller black holes.  For very large black holes, the thermal fluctuation does not effect significantly.
 The negative cosmological constant is assumed as the positive thermodynamic pressure. Once the leading-order corrected entropy, the temperature, horizon angular velocity,  angular momentum  charge and pressure are known, we can easily  
 derive  the first-order corrected physical mass of the system. 
 We find that there exists a critical point for physical mass which increases as long as charge of the black hole increases.
The mass is a  decreasing function before the critical   horizon radius and thereafter it is an
increasing function. The correction terms with negative correction parameter increases and decreases the
values of physical mass before and after the critical horizon radius respectively.
 Once the expression of corrected physical mass and pressure is known, it is matter of calculation to 
 derive the geometrical volume for a charged rotating AdS black hole. We estimate the correction to internal energy also. We find that internal energy is a decreasing function of horizon radius for smaller black holes. There exists a critical
value for internal energy which increases when charge increase of the black hole. The correction term
due to thermal fluctuation with negative correction parameter  increases and decreases the value of
internal energy before and after critical point respectively.  We further calculated the
corrected value of Helmholtz and Gibbs free energies of the system due to thermal fluctuations. 
    We further study the Van der waals fluid duality to charged rotating black holes 
and see the typical behavior of the $P-v$ diagram corresponding to the van der Waals fluid.
 The stability of such black holes is also checked. We find that  the phase transition occurs for such black holes. The specific heat is negative for small black
holes which suggests that small black holes are thermodynamically unstable. In fact, the
specific heat is always positive for larger black holes which means these black holes are in stable phase. Some stability also occurs for the small black holes due to the correction term with positive correction parameter. However, the correction term with negative correction parameter makes small black holes more unstable.
  
  The paper is organized as follows.  In section \ref{black}, we first recapitulate the   
metric and thermodynamical quantities of  a charged Kerr-AdS black hole. Then, we study the
first-order correction to thermodynamics of such black hole. Further, we analyse the
duality between black hole and Van der Waals fluid. We study  the stability of such black holes also. Finally, we conclude the results with future directions in section \ref{con}.
\section{\label{black}A charged rotating black holes in AdS space} 
The charged rotating (Kerr) 
black holes in $d=4$  AdS space were constructed  by Carter \cite{car} and the metric reads,
\begin{eqnarray}
ds^2=-\frac{\Delta}{\rho^2}\left[dt -\frac{a}{\Sigma}\sin^2\theta d\phi \right]^2+\frac{\rho^2dr^2}{\Delta}+\frac{\rho^2d\theta^2}{\Delta_\theta}+\frac{\Delta_\theta \sin^2\theta}{\rho^2}\left[adt-\frac{r^2+a^2}{\Sigma}d\phi \right]^2,
\end{eqnarray}
where
\begin{eqnarray}
&&\Delta=(r^2+a^2)\left(1+\frac{r^2}{l^2}\right)-2mr+q^2,\ \ \ \ \Delta_\theta =1-\frac{a^2}{l^2}\cos^2\theta,\nonumber\\
&&\rho^2=r^2+a^2\cos^2\theta,\ \ \ \ \ \ \Sigma =1-\frac{a^2}{l^2}.
\end{eqnarray}
Here $l$ is the AdS radius and $a$ is the rotation parameter.
The associated electromagnetic field is due to the
potential one-form $A_\mu=-\frac{q r}{\rho}\left(dt-\frac{a\sin^2\theta}{\Sigma}d\phi\right)$, 
where the parameter $q$ is related to the charge of black hole $Q=\frac{q}{\Sigma}$.
 This metric satisfies  the vacuum Einstein equations $R_{\mu\nu}=-3 {g_{\mu\nu}}/{l^2}$. 
The outer horizon is located at the radius $r=r_+$
 which is determined by the equation
\begin{eqnarray}
2m=\frac{(r_+^2+a^2)(r_+^2+l^2)}{r_+l^2}+\frac{q^2}{r_+}.
\end{eqnarray}
The   entropy of the system in thermal equilibrium can be derived in Planck's unit as
\begin{eqnarray}
S_0=\frac{A}{4}=\frac{\pi  (r^2_++a^2)}{\Sigma}.\label{so}
\end{eqnarray}
The Hawking temperature is given by \cite{gib}
\begin{eqnarray}
T=\frac{\kappa}{2\pi}=\frac{3r^4_++r_+^2l^2+a^2r_+^2-l^2(a^2+q^2)}{4\pi l^2r_+(r_+^2+a^2)},
\label{t}
\end{eqnarray}
where $\kappa$ refers to the surface gravity.
These black holes have the
following roots for the temperature:
\begin{eqnarray}
r|_T=\pm \left[\frac{-a^2-l^2\pm \sqrt{(a^2+l^2)^2 +12l^2(a^2+q^2)}}{6}\right]^{1/2}.
\end{eqnarray}
It is evident that the values in the square bracket are both  positive and negative.  
So, the existence of a real valued roots for the temperature is
limited to the following condition $-a^2-l^2\pm \sqrt{(a^2+l^2)^2 +12l^2(a^2+q^2)}>0$.
Here, we see that for Kerr-AdS black holes two real valued root exist 
for the temperature.

The other thermodynamic quantities  (in Planck units) are given by \cite{gib}
\begin{eqnarray}
&&\Omega_H =-\left.\frac{g_{t\phi}}{g_{\phi\phi}}\right|_{r=r_+}=\frac{a(l^2+r_+^2)}{l^2(a^2+r_+^2)},\ \ \ J=\frac{a(r_+^2+a^2)(1+r_+^2l^{-2})+aq^2}{2r_+\Sigma^2}, \nonumber\\
&&  M_0=\frac{(r_+^2+a^2)(1+r_+^2l^{-2})+q^2}{2r_+\Sigma^2},\ \  \ \Phi=\frac{qr_+}{r_+^2+a^2},
\end{eqnarray}
where $\Omega_H$, $J$, $M_0$  and $\Phi$ denote the horizon angular velocity  measured relative to a frame that is non-rotating at infinity, angular momentum, the  physical mass (or energy)  and the electric potential measured at infinity with respect to the horizon of the black hole, respectively.
\subsection{ Thermodynamics with first-order correction}
 In this section, we study the effects of quantum corrections to the thermodynamics of
 rotating black hole upto first-order. We assume the system of black hole follows the 
  canonical ensemble.   In this regard, let us start by
defining the density of states with fixed energy   as \cite{boh,rk}
\begin{eqnarray}
\rho =\frac{1}{2\pi i}\int_{c-i\infty}^{c+i\infty}e^{{\cal S}(\beta)}d\beta,\label{rho}
\end{eqnarray}
where ${\cal S}(\beta)$ is the exact entropy as a function of temperature $T=1/\beta$.
The above complex integral can be solved by the method of steepest descent around the saddle point $\beta_0$. 
Now, by expanding the exact entropy around the saddle point $\beta=\beta_0$, we have
\begin{eqnarray}
{\cal S}=S_0+\frac{1}{2}(\beta-\beta_0)^2 S_0''+ \mbox{(higher order terms)},\label{s}
\end{eqnarray}
where $S_0={\cal S}(\beta_0)$ refers the zeroth-order entropy and $S_0'':=\left(\frac{\partial^2 {\cal S}(\beta)}{\partial \beta^2}\right)_{\beta=\beta_0}$.
Inserting this value of ${\cal S}$ given in (\ref{s})  to (\ref{rho}), and performing integral
we get \cite{l1}
\begin{eqnarray}
\rho=\frac{e^{S_0}}{\sqrt{2\pi S_0''}}.
\end{eqnarray}
This leads to  
microcanonical entropy
\begin{eqnarray}
S=S_0+\alpha\log S_0''+.... .
\end{eqnarray}
where   $\alpha$ is a correction coefficient. 
By considering most general form  of ${\cal S}$, the   $S_0''$ can be determined.
Therefore, the  generic expression for leading order corrections to
Bekenstein-Hawking formula is  given by \cite{l1}
\begin{eqnarray}
S &=& S_{0} + \alpha \ln (S_{0} T_H^{2}),\nonumber\\
&=&  S_{0} + \alpha \ln  S_{0}+  2\alpha \ln T_H. \label{correctedS}
\end{eqnarray}
The first law of thermodynamics
for black holes and Smarr-Gibbs-Duhem relation read,
\begin{eqnarray}
dM&=&TdS+\Omega_H dJ+VdP+\Phi dQ,\nonumber\\
M&=&2TS+2\Omega J+ \Phi Q-2PV.\label{m}
\end{eqnarray}
Here, the negative cosmological constant  can be interpreted  as
the positive thermodynamic pressure 
\begin{eqnarray}
P=-\frac{\Lambda}{8\pi}=\frac{3}{8\pi l^2}.\label{pre}
\end{eqnarray} 
Now, plugging the values of $T$ and $S_0$ from (\ref{t}) and (\ref{so})   respectively  to  the relation (\ref{correctedS}), we get the leading-order corrected entropy is given by
 \begin{eqnarray}\label{cs}
 S=\frac{\pi  (r^2_++a^2)}{\Sigma}+\alpha\log \frac{\pi  (r^2_++a^2)}{\Sigma} + 2 \alpha\log  \left[\frac{3r^4_++r_+^2l^2+a^2r_+^2-l^2(a^2+q^2)}{4\pi l^2r_+(r_+^2+a^2)} \right].
 \end{eqnarray}

  \begin{figure}[htb]
 \begin{center}
 $
 \begin{array}{cc}
\includegraphics[width=70 mm]{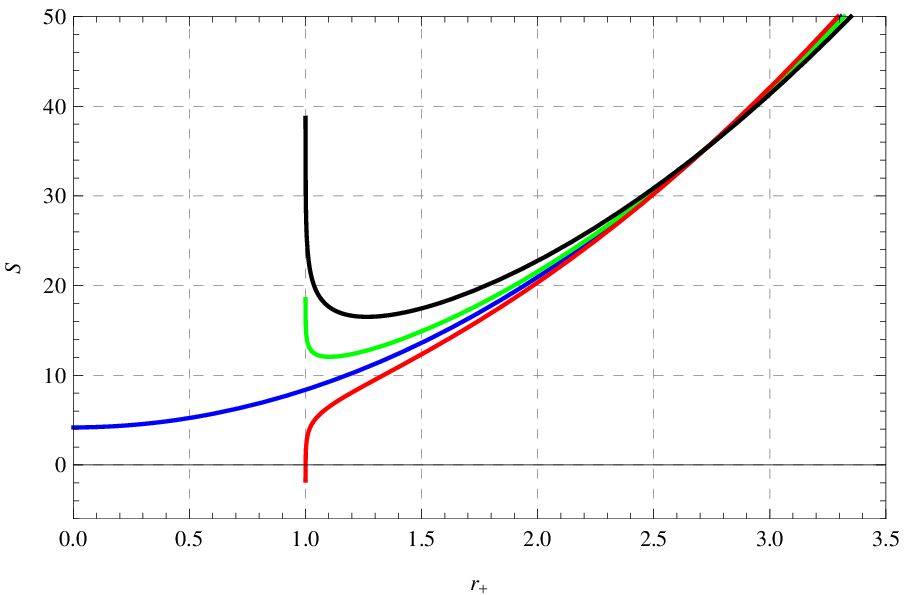}    \ \ \ \ & \includegraphics[width=70 mm]{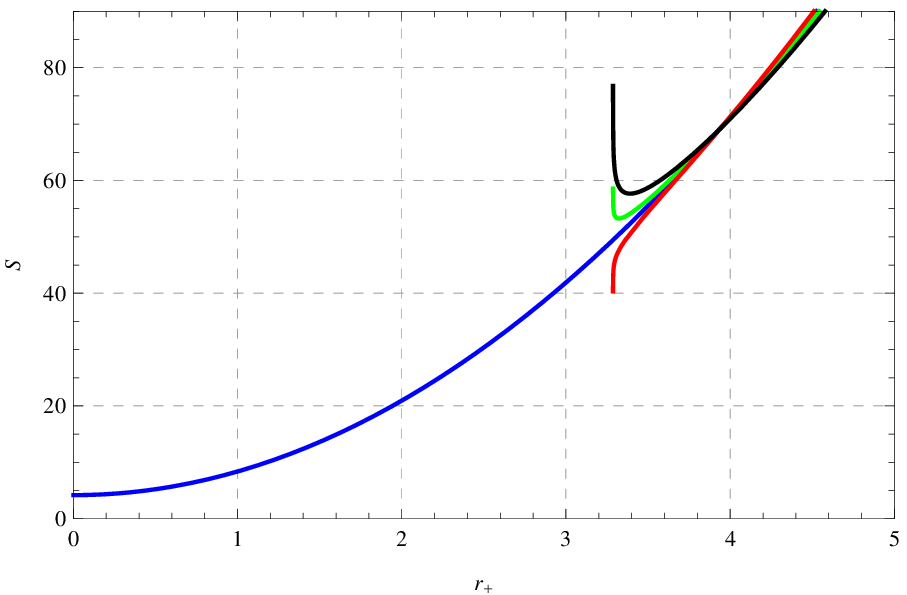} 
 \end{array}$
  \end{center}
 \caption{Left: Entropy vs. the black hole horizon for $a=1, q=1$ and $l=2$. Right: Entropy vs. the black hole horizon for $a=1, q=10$ and $l=2$. Here, $\alpha=0$ denoted by blue line, $\alpha=-0.5$ denoted by green  line, $\alpha=0.5$ denoted by red line, and $\alpha=-1.5$ denoted by black line.}
 \label{fig1}
\end{figure}
The effects of thermal fluctuation on entropy can be seen in Fig. \ref{fig1}. We see that
entropy is an increasing function for large black holes as well as if the system is in equilibrium. Due to thermal fluctuation, the entropy does not exist for smaller black holes.
As charge increases the entropy does not occur for even larger black holes.
For very large black holes, the  thermal fluctuation does not play an important role.

 From the first expression of  (\ref{m}), the corrected physical mass for charged rotating black hole is calculated by
\begin{eqnarray}
M&=& \frac{(r_+^2+a^2)(1+r_+^2l^{-2})+q^2}{2r_+\Sigma^2}+\alpha\frac{6a^4r_++6a^2r_+^3+2a^4r_+-2a^2l^2r_+-l^2q^2r_+}{4\pi l^2a^2(a^2+r_+^2)}\nonumber\\
&-&\alpha\frac{(8a^4+l^2q^2)}{4\pi l^2 a^3}\tan^{-1}\left[\frac{r_+}{a}\right]+\alpha \frac{3r^4_++r_+^2l^2+a^2r_+^2-l^2(a^2+q^2)}{2\pi l^2r_+(r_+^2+a^2)}.
 \end{eqnarray}
  \begin{figure}[htb]
 \begin{center}
 $
 \begin{array}{cc}
\includegraphics[width=70 mm]{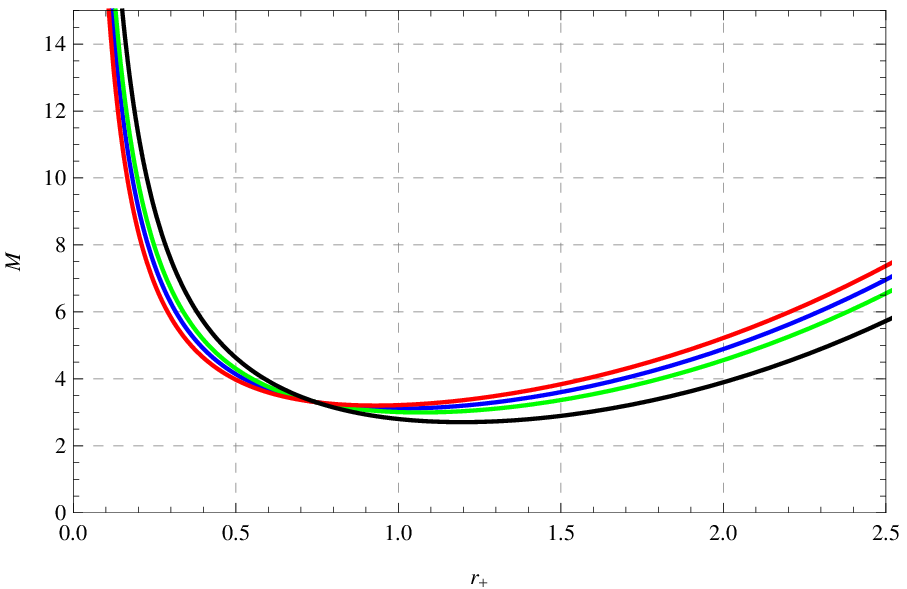}    \ \ \ \ & \includegraphics[width=70 mm]{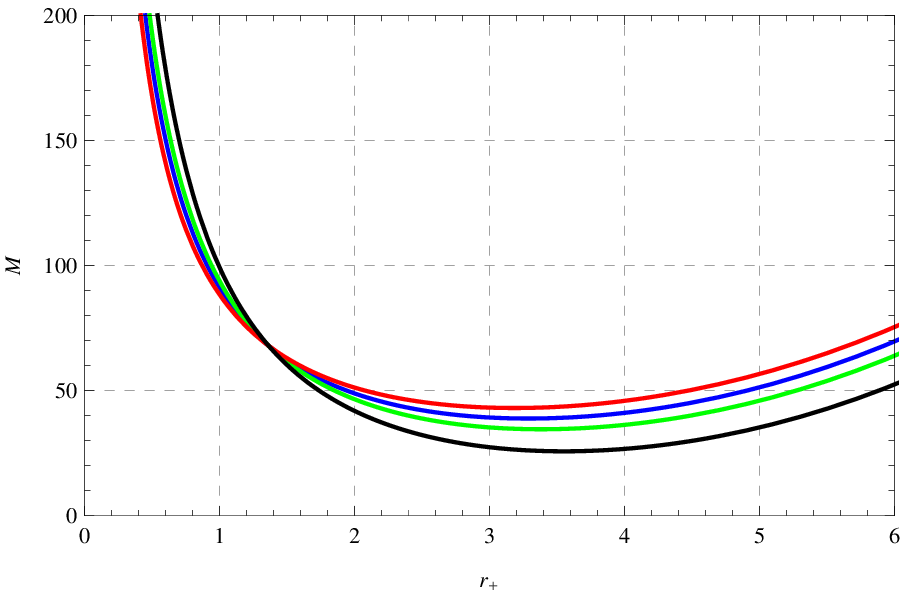} 
 \end{array}$
  \end{center}
 \caption{Left: Physical mass vs. the black hole horizon for $a=1, q=1$ and $l=2$. Right: Physical mass vs. the black hole horizon for $a=1, q=10$ and $l=2$. Here, $\alpha=0$ denoted by blue line, $\alpha=-0.5$ denoted by green  line, $\alpha=0.5$ denoted by red line, and $\alpha=-1.5$ denoted by black line.}
 \label{fig2}
\end{figure}
We plot Fig. \ref{fig2} to see the behavior of physical mass with respect to horizon radius.
Here, one can see that there exists a critical point for physical mass which increases as long as charge of the black hole increases. The mass is an decreasing function before this critical value of horizon radius and thereafter it is an increasing function. The correction terms
with negative correction parameter increases and decreases the values of physical mass 
before and after the critical horizon radius respectively.

The geometrical volume for charged rotating AdS black hole in $d=4$ is calculated by  
  \begin{eqnarray}
 V= \left(\frac{\partial M_0}{\partial P}\right)_{S_0}=\frac{4}{3} \frac{\pi r_+(r_+^2+a^2)}{\Sigma^2}.\label{v}
 \end{eqnarray}
Using definition, $U=M-PV$,  the internal energy is calculated by  
\begin{eqnarray}
U&=& \frac{(r_+^2+a^2)(1+r_+^2l^{-2})+q^2}{2r_+\Sigma^2}-\frac{r_+(r^2+a^2)}{2l^2\Sigma^2}+\alpha\frac{6a^4r_++6a^2r_+^3+2a^4r_+-2a^2l^2r_+-l^2q^2r_+}{4\pi l^2a^2(a^2+r_+^2)}\nonumber\\
&-&\alpha\frac{(8a^4+l^2q^2)}{4\pi l^2 a^3}\tan^{-1}\left[\frac{r_+}{a}\right]+\alpha \frac{3r^4_++r_+^2l^2+a^2r_+^2-l^2(a^2+q^2)}{2\pi l^2r_+(r_+^2+a^2)}.
 \end{eqnarray}
 \begin{figure}[htb]
 \begin{center}
 $
 \begin{array}{cc}
\includegraphics[width=70 mm]{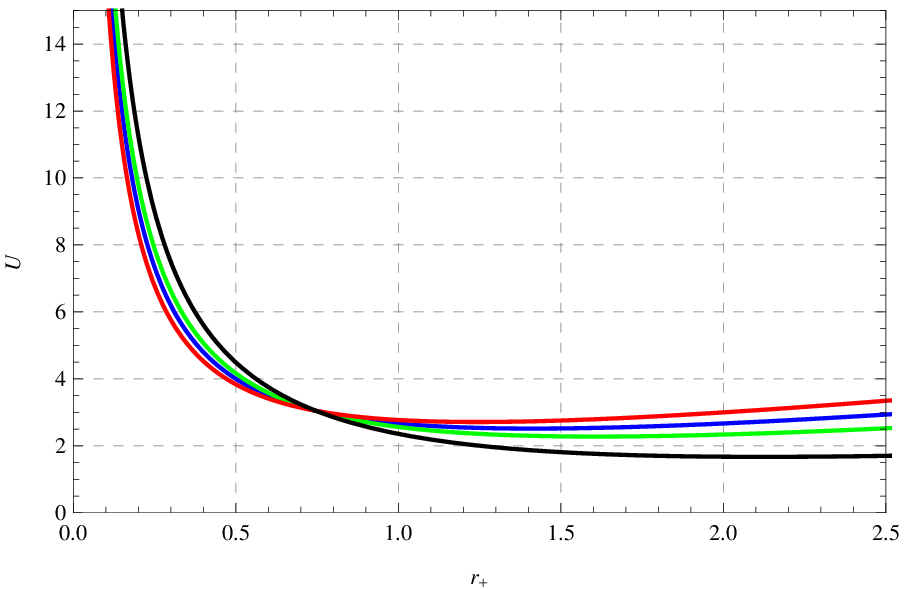}    \ \ \ \ & \includegraphics[width=70 mm]{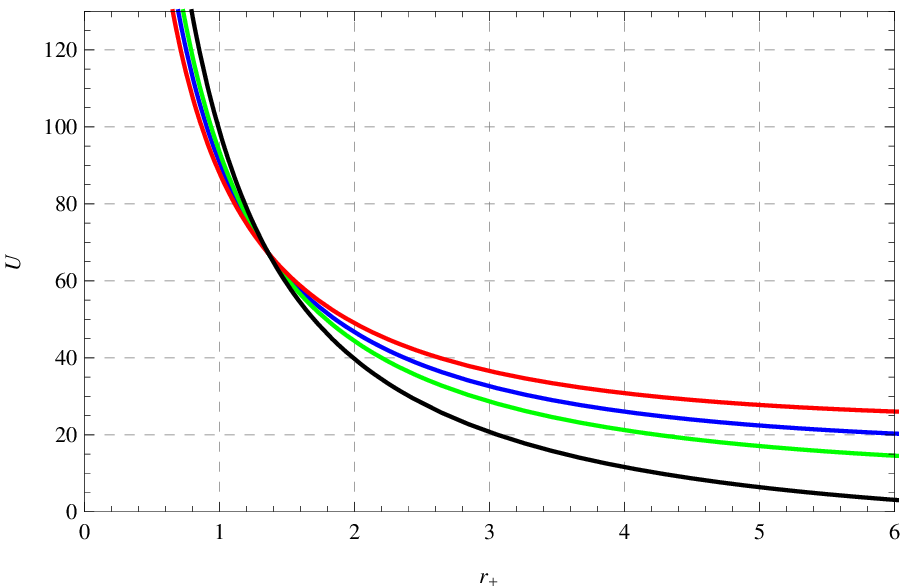} 
 \end{array}$
  \end{center}
 \caption{Left: Internal energy vs. the black hole horizon for $a=1, q=1$ and $l=2$. Right: Internal energy  vs. the black hole horizon for $a=1, q=10$ and $l=2$. Here, $\alpha=0$ denoted by blue line, $\alpha=-0.5$ denoted by green  line, $\alpha=0.5$ denoted by red line, and $\alpha=-1.5$ denoted by black line.}
 \label{fig20}
\end{figure}
We behavior of internal energy with respect to horizon radius can be seen in Fig. \ref{fig20}.
We see that internal energy is a decreasing function of horizon radius for smaller black holes. There exists a critical value for internal energy which increases when charge increase
of the black hole. The correction term due to thermal fluctuation with negative correction parameter $\alpha$ increases and decreases the value of  internal energy before and after critical point respectively. However, The correction term due to thermal fluctuation with positive correction parameter $\alpha$ decreases and increases the value of  internal energy before and after critical point respectively.

 Once the expressions for Hawking temperature,  entropy and physical mass are known, it is matter of
 calculation to evaluate the Helmholtz  free energy:  $F=M-TS$.
 Hence, from (\ref{cs}) and (\ref{t}), the first-order corrected Helmholtz  free energy is calculated by
\begin{eqnarray}
F&=& \frac{1}{4r_+\Sigma^2}\left(r_+^2 -r_+^4l^{-2}+3a^2 +2a^2r_+^2l^{-2}+3q^2  +3a^2l^{-4}r_+^4+a^4l^{-4}r_+^2-a^4l^{-2}-a^2q^2l^{-2} \right)\nonumber\\
 &+&  \alpha\frac{6a^4r_++6a^2r_+^3+2a^4r_+-2a^2l^2r_+-l^2q^2r_+}{4\pi l^2a^2(a^2+r_+^2)}-\alpha\frac{(8a^4+l^2q^2)}{4\pi l^2 a^3}\tan^{-1}\left[\frac{r_+}{a}\right]\nonumber\\
&+& \alpha \frac{3r^4_++r_+^2l^2+a^2r_+^2-l^2(a^2+q^2)}{2\pi l^2r_+(r_+^2+a^2)}-\alpha \frac{3r^4_++r_+^2l^2+a^2r_+^2-l^2(a^2+q^2)}{4\pi l^2r_+(r_+^2+a^2)}\log \left[\frac{\pi  (r^2_++a^2)}{\Sigma} \right]\nonumber\\
&-&2\alpha \frac{3r^4_++r_+^2l^2+a^2r_+^2-l^2(a^2+q^2)}{4\pi l^2r_+(r_+^2+a^2)}\log \frac{3r^4_++r_+^2l^2+a^2r_+^2-l^2(a^2+q^2)}{4\pi l^2r_+(r_+^2+a^2)}. \label{f}
 \end{eqnarray}
 \begin{figure}[htb]
 \begin{center}$
 \begin{array}{cc}
\includegraphics[width=70 mm]{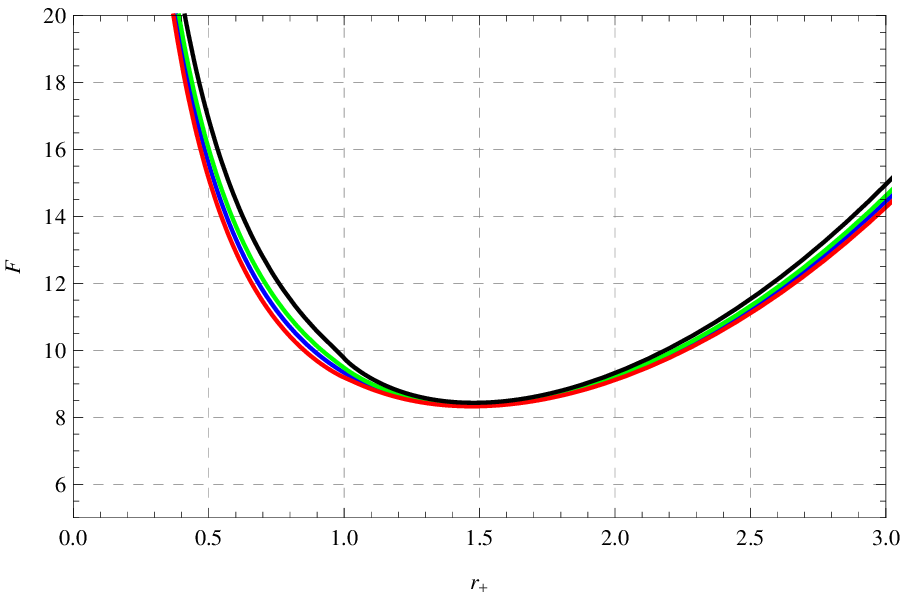}  \ \ \ \ & \includegraphics[width=70 mm]{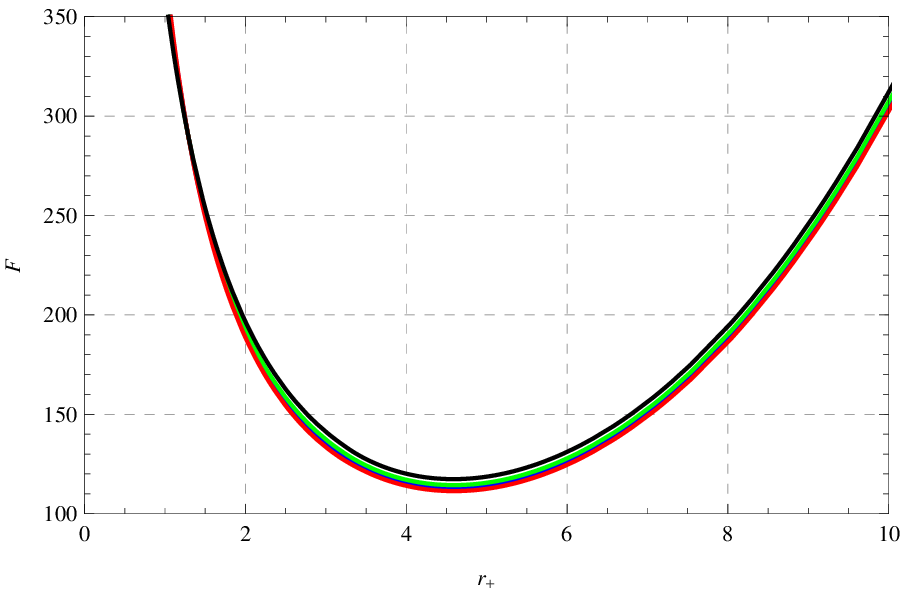} 
 \end{array}$
 \end{center}
\caption{Left: Helmholtz free energy   vs. the black hole horizon for $a=1, q=1$ and $l=2$. Right: Helmholtz free energy  vs. the black hole horizon for $a=1, q=10$ and $l=2$. Here, $\alpha=0$ denoted by blue line, $\alpha=-0.5$ denoted by green  line, $\alpha=0.5$ denoted by red line, and $\alpha=-1.5$ denoted by black line.}
 \label{fig3}
\end{figure}
The effects of leading-order correction to  Helmholtz free energy can be seen in Figure  \ref{fig3}. It can be seen that the behavior of Helmholtz free energy is decreasing for small black holes and increasing for larger black holes. The value of Helmholtz free energy is always positive. The higher charge increases the value of Helmholtz free energy.
Unlike the internal energy, the correction term with negative $\alpha$ increases the value of 
Helmholtz free energy always (before and after critical value).

Since mass of the black hole $M$ is interpreted as the enthalpy, we have the following
expression for the Gibbs free energy : $G=M-TS-\Phi Q$, which can be used to calculate the maximum of reversible work that may be performed by the  system
of rotating black hole. This is calculated by
 \begin{eqnarray}
 G&=& \frac{1}{4r_+\Sigma^2}\left(r_+^2 -r_+^4l^{-2}+3a^2 +2a^2r_+^2l^{-2}+3q^2  +3a^2l^{-4}r_+^4+a^4l^{-4}r_+^2-a^4l^{-2}-a^2q^2l^{-2} \right)\nonumber\\
 &-& \frac{q^2r_+}{\Sigma(r_+^2+a^2)}+\alpha\frac{6a^4r_++6a^2r_+^3+2a^4r_+-2a^2l^2r_+-l^2q^2r_+}{4\pi l^2a^2(a^2+r_+^2)}-\alpha\frac{(8a^4+l^2q^2)}{4\pi l^2 a^3}\tan^{-1}\left[\frac{r_+}{a}\right]\nonumber\\
&+& \alpha \frac{3r^4_++r_+^2l^2+a^2r_+^2-l^2(a^2+q^2)}{2\pi l^2r_+(r_+^2+a^2)}-\alpha \frac{3r^4_++r_+^2l^2+a^2r_+^2-l^2(a^2+q^2)}{4\pi l^2r_+(r_+^2+a^2)}\log \left[\frac{\pi  (r^2_++a^2)}{\Sigma} \right]\nonumber\\
&-&2\alpha \frac{3r^4_++r_+^2l^2+a^2r_+^2-l^2(a^2+q^2)}{4\pi l^2r_+(r_+^2+a^2)}\log \frac{3r^4_++r_+^2l^2+a^2r_+^2-l^2(a^2+q^2)}{4\pi l^2r_+(r_+^2+a^2)}.
 \end{eqnarray}
 \begin{figure}[htb]
 \begin{center}$
 \begin{array}{cc}
\includegraphics[width=70 mm]{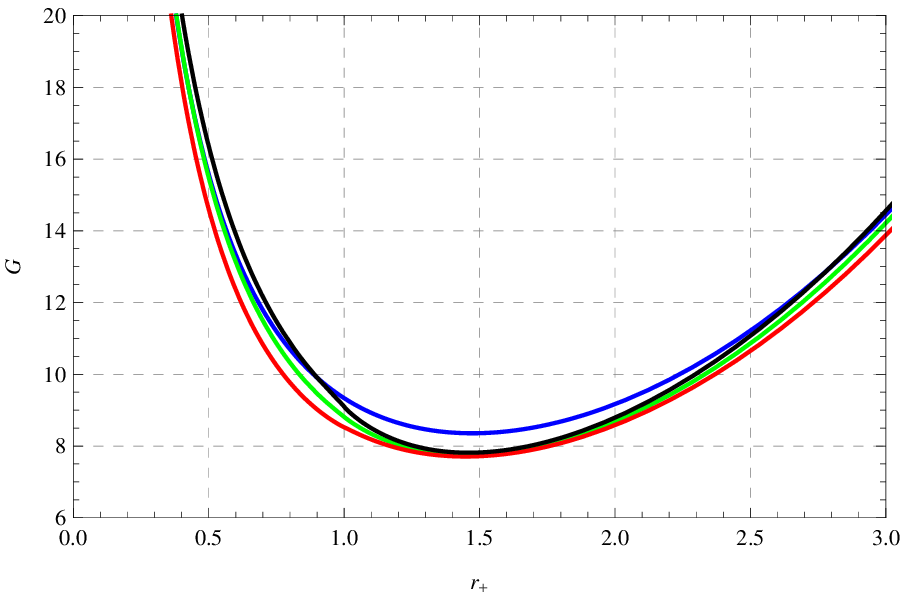}    \ \ \ \ & \includegraphics[width=70 mm]{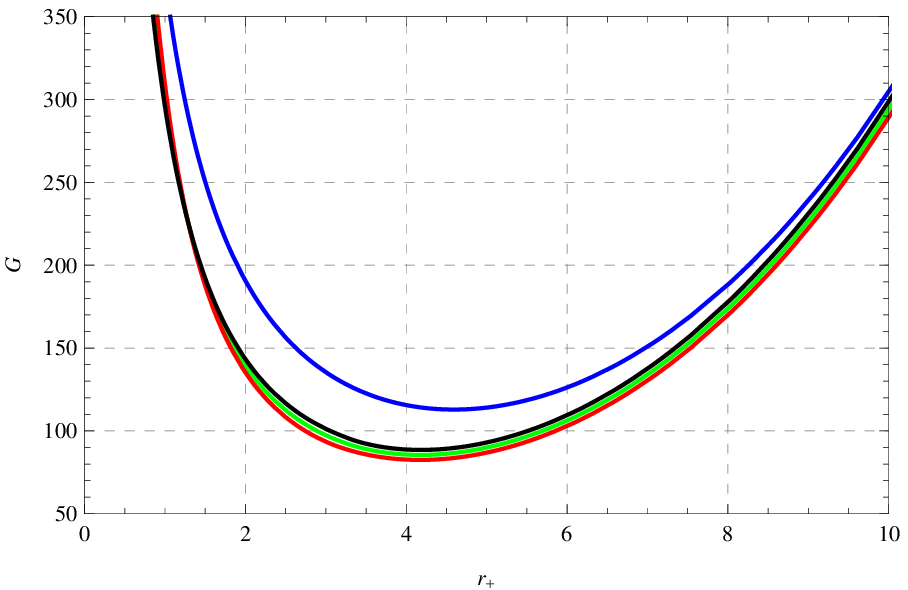} 
 \end{array}$
 \end{center}
\caption{Left: Gibbs free energy   vs. the black hole horizon for $a=1, q=1$ and $l=2$. Right: Gibbs free energy  vs. the black hole horizon for $a=1, q=10$ and $l=2$. Here, $\alpha=0$ denoted by blue line, $\alpha=-0.5$ denoted by green  line, $\alpha=0.5$ denoted by red line, and $\alpha=-1.5$ denoted by black line.}
 \label{fig4}
\end{figure}
The graphical analysis of Gibbs free energy for the rotating black hole  can be seen in Fig.  \ref{fig4}. Here, we observe that correction terms decreases the value of  Gibbs free energy
for smaller black hole. As long as charge of black hole increases, the correction terms due to thermal fluctuation decrease the Gibbs free energy. For large black hole, the correction terms 
do not effect the Gibbs free energy much.  
\subsection{Van der Waals fluid duality}
We realize that the rotating black holes behave as Van der Walls fluids. The Van der Walls
equation describes the behavior of real fluids and modifies the ideal gas equation of states as
\begin{eqnarray}
\left(P+\frac{\mathfrak{a}}{v^2}\right)(v-\mathfrak{b})=T \Rightarrow P=\frac{T}{v-\mathfrak{b}}-
\frac{\mathfrak{a}}{v^2},\label{pp}
\end{eqnarray}
where $v=V/N$ is the specific  volume of the fluid. Here $N=A/l_P^2$ denotes the number of degrees of freedom associated with the black hole horizon (plank length) $l_P=\sqrt{G\bar h/c^2}$. The  constant $\mathfrak{a} > 0$ is a measure of the interaction
between  the molecules of a given fluid while the constant $\mathfrak{b} > 0$
refers to the nonzero size of them.  
 The expression for specific volume 
\cite{mann} is
given by
\begin{eqnarray}
v= 6\frac{V}{A}=2\frac{r_+}{\Sigma}.\label{vv}
\end{eqnarray}
as in the geometric units ($G=\hbar=c=k_B=1$) Planck length is  unit.
Utilizing the relations   (\ref{pre}) and (\ref{vv}), we can plot graph between pressure and specific volume.
\begin{figure}[htb]
 \begin{center}$
 \begin{array}{cc }
\includegraphics[width=70 mm]{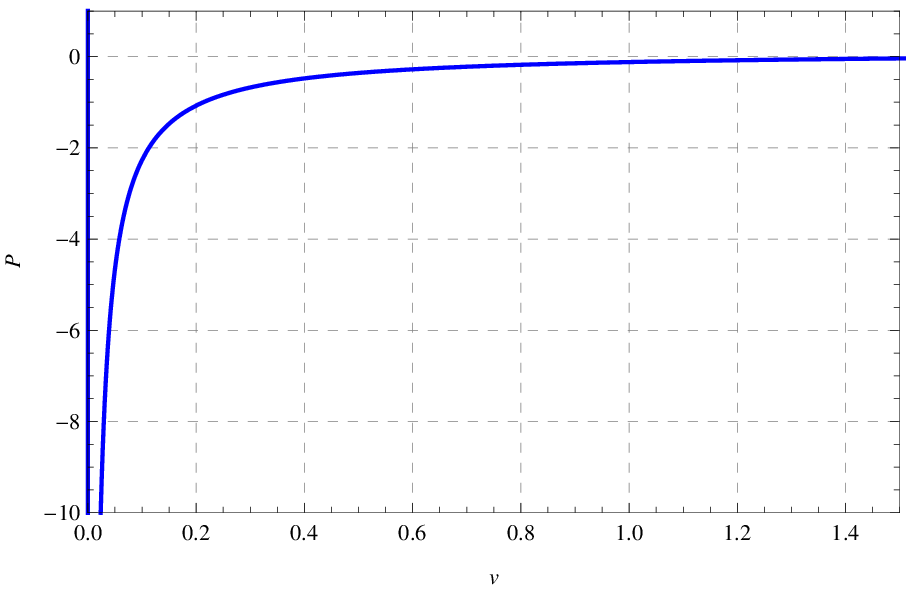}  \ \ \ \ &  \includegraphics[width=70 mm]{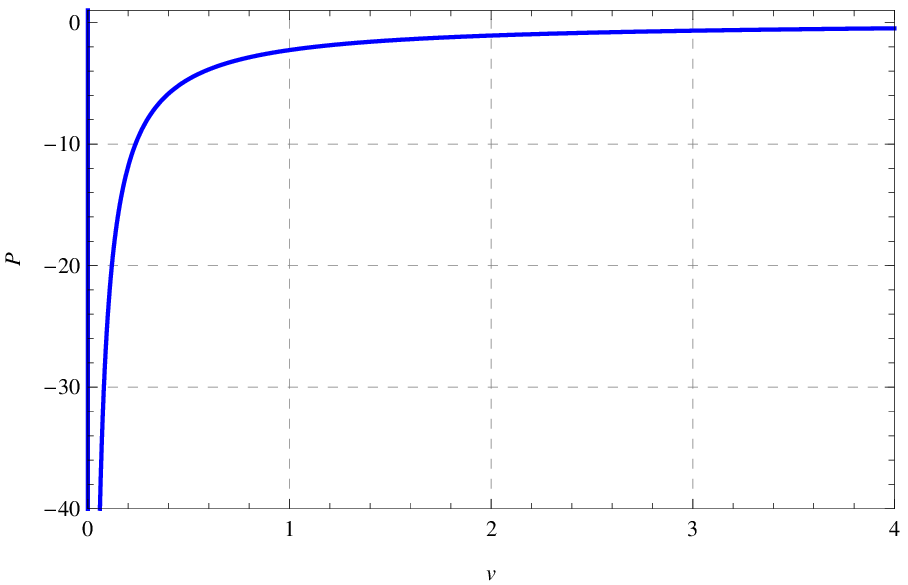}
 \end{array}$
 \end{center}
\caption{Left: Pressure   vs. the specific volume for $a=1$ and $r_+=1$. Right: Pressure   vs. the specific volume for $a=1$ and $r_+=10$}
 \label{fig6}
\end{figure}
From the fig. (\ref{fig6}), the typical behavior of the  $P-v$ diagram corresponding
to the van der Waals fluid can be seen. The pressure is negative for small black holes with respect to specific volume. 
To study critical phenomena and van der Waals behavior,
we compute the following equation of state:
\begin{eqnarray}
P=\frac{3(r_+^2+a^2)T}{2r_+(3r_+^2+a^2)}+\frac{3(a^2+q^2)}{8\pi r_+^2(3r_+^2+a^2)}-\frac{3}{8\pi(3r_+^2+a^2)(r_+^2+a^2)}.
\end{eqnarray}

The critical points can be obtained in this case  from conditions  $\left(\frac{\partial P}{\partial r_+}\right)_T=0$ and $\left(\frac{\partial^2 P}{\partial r_+^2}\right)_T=0$.
 
\subsection{Stability}
The stability  and phase transition of a thermal system can be studied by analyzing 
the specific heat of the system.
The specific heat at constant charge can be calculated from following relation:
\begin{eqnarray}
C_Q=T\left(\frac{\partial S}{\partial T}\right)_Q.
\end{eqnarray}
From the above definition,  the   first-order corrected specific heat for a charge rotating black hole is derived by
\begin{eqnarray}
C_Q&=&\frac{2 \pi  r_+^2 \left(a^2+r_+^2\right) \left[a^2 \left(r_+^2-l^2\right)+l^2 (Q \Sigma +r) (r_+-Q \Sigma )+3 r_+^4\right]}{\Sigma  \left[a^4 \left(l^2+r_+^2\right)+a^2 \left(l^2 \left(Q^2 \Sigma ^2+4 r_+^2\right)+8 r^4\right)-l^2 \left(r_+^4-3 Q^2 r_+^2 \Sigma ^2\right)+3 r_+^6\right]}\nonumber\\
&+&2\alpha\frac{  r_+^2   \left[a^2 \left(r_+^2-l^2\right)+l^2 (Q \Sigma +r) (r_+-Q \Sigma )+3 r_+^4\right]}{  \left[a^4 \left(l^2+r_+^2\right)+a^2 \left(l^2 \left(Q^2 \Sigma ^2+4 r_+^2\right)+8 r_+^4\right)-l^2 \left(r_+^4-3 Q^2 r_+^2 \Sigma ^2\right)+3 r_+^6\right]}\nonumber\\
&+&2\alpha.
\end{eqnarray}
\begin{figure}[htb]
 \begin{center}$
 \begin{array}{cc }
\includegraphics[width=70 mm]{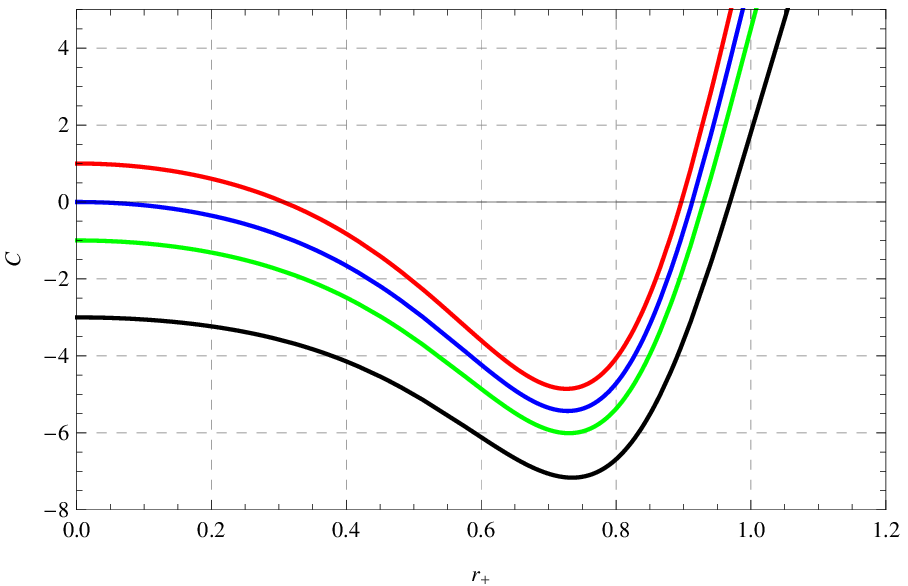}  \ \ \ \ &  \includegraphics[width=70 mm]{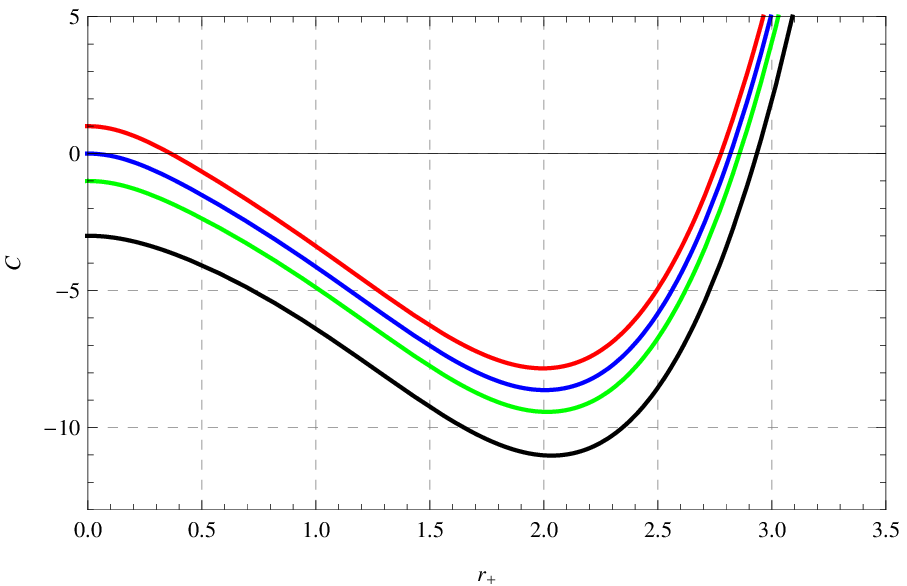}
 \end{array}$
 \end{center}
\caption{Left: Specific heat   vs. the black hole horizon for $a=1, q=1$ and $l=2$. Right: Specific heat  vs. the black hole horizon for $a=1, q=10$ and $l=2$. Here, $\alpha=0$ denoted by blue line, $\alpha=-0.5$ denoted by green  line, $\alpha=0.5$ denoted by red line, and $\alpha=-1.5$ denoted by black line.}
 \label{fig7}
\end{figure}
In fig. (\ref{fig7}), we see that the behavior of specific heat of a charged rotating black holes. This diagram suggests that 
the   phase transition  occurs for such black hole.
The specific heat   is negative  for small  black holes which suggests that small  black 
holes are thermodynamically unstable. On the other hand, the specific heat is always positive for larger black holes which means these black holes are in stable phase. 
Some stability also occurs for the small black holes due to the  correction term  with positive correction parameter. However, the correction term with negative $\alpha$ makes
small black holes more unstable.

The divergences occurs at $l=a$ and
\begin{eqnarray}
a^4 \left(l^2+r_+^2\right)+a^2  l^2 \left(Q^2 \Sigma ^2+4 r_+^2\right)+8a^2 r_+^4 -l^2 \left(r_+^4-3 Q^2 r_+^2 \Sigma ^2\right)+3 r_+^6=0.
\end{eqnarray}
This can be solved easily and the real solutions are
\begin{eqnarray}
r=\pm \left(\frac{\sqrt[3]{2}(55   a^4-52   a^2 l^2+  l^4-27  l^2 Q^2 \Sigma ^2)}{9 \sqrt[3]{\sqrt{\beta^2+\gamma}+\beta}}  +\frac{l^2-8 a^2}{9}+\frac{\sqrt[3]{\sqrt{\beta^2 +\gamma}+\beta}}{9 \sqrt[3]{2}}\right)^{1/2},\label{r}
\end{eqnarray}
where  $\beta$ and  $\gamma$ are
\begin{eqnarray}\beta &=& -808 a^6+978 a^4 l^2-156 a^2 l^4+405 a^2 l^2 Q^2 \Sigma ^2+2 l^6-81 l^4 Q^2 \Sigma ^2,\nonumber\\
 \gamma &=&  4 \left(-55 a^4+52 a^2 l^2-l^4+27 l^2 Q^2 \Sigma ^2\right)^3.
\end{eqnarray}
 Here, we assumed that the overall value of expressions  inside the bracket in (\ref{r}) is positive. 

\section{\label{con}Concluding remarks}
In this paper, we have considered a charged rotating AdS black hole with their  
thermodynamical quantities when system is in equilibrium and studied the effects of leading-order corrections
stemming from thermal fluctuations of charged rotating black holes on equations of states.  
 In particular, we have  derived the
 Hawking temperature and  logarithmic correction to the entropy first which leads to correction to other 
 thermodynamical quantities.  
 We found  that entropy is an increasing
function of horizon radius for large black holes when system is in equilibrium. Due to thermal fluctuation, the entropy does not depend on the horizon radius  for smaller black holes. We found that for very large black holes, the thermal fluctuation does not effect significantly.
In our analysis, the negative cosmological constant is assumed as the positive thermodynamic pressure. After having expressions of leading-order corrected entropy, the temperature, horizon angular velocity,  angular momentum  charge and pressure, we have 
 derived  the first-order corrected physical mass of the system from the first-law of thermodynamics. 
 We observed  that there exists a critical value physical mass which increases as long as charge of the black hole increases.
From the plot, one can see that the mass is a  decreasing function before the critical  point and thereafter it is an
increasing function. The correction terms with negative correction parameter increases 
values of physical mass before  the critical point and decreases the physical mass after 
the critical point. 
Utilizing the standard relation between  physical mass and geometrical volume,  we have calculated the geometrical volume of the charged rotating AdS black holes. We have 
evaluated  the leading-order correction to internal energy also. In this regard, we have found that internal energy is a decreasing function of horizon radius for smaller black holes. If fact, there exists a critical
value for internal energy as well which increases when charge of the black hole increases. The correction term with negative correction parameter 
originated from the thermal fluctuation  increases  the value of
internal energy before  the critical point and decreases tits value after the critical point.  Moreover, we have calculated the
corrected value of Helmholtz   free energies of the system due to thermal fluctuations. 
It is observed  that the  Helmholtz free energy is decreasing function of horizon radius for small black holes and increasing function for larger black holes. The value of Helmholtz free energy is found always positive. The higher value of charge increases the value of Helmholtz free energy.
Unlike the internal energy, the correction term with negative $\alpha$ always increases the value of 
Helmholtz free energy. We have estimated the correction to Gibbs free energy as well.

    We have studied the Van der waals fluid duality to charged rotating black holes 
and analyzed the typical behavior of the $P-v$ diagram corresponding to the van der Waals fluid.
 The stability of such black holes is also checked. We have found that  a phase transition occurs for such black holes. For small black
holes, the specific heat takes negative values which suggests that small black holes are thermodynamically unstable phase. However, the
specific heat is found always positive for larger black holes which means that these black holes are in stable phase. The stability also occurs for the small black holes in case of 
thermal fluctuations also when correction parameter takes positive values. The
 correction term with negative correction parameter originated from thermal fluctuations  introduces more instabilities to small black holes.

 \end{document}